\input harvmac
%\draftmode
\let\includefigures=\iftrue
\let\useblackboard==\iftrue
\newfam\black

%Figure Stuff
\includefigures
\message{If you do not have epsf.tex (to include figures),}
\message{change the option at the top of the tex file.}
\input epsf
\def\figin{\epsfcheck\figin}\def\figins{\epsfcheck\figins}
\def\epsfcheck{\ifx\epsfbox\UnDeFiNeD
\message{(NO epsf.tex, FIGURES WILL BE IGNORED)}
\gdef\figin##1{\vskip2in}\gdef\figins##1{\hskip.5in}% blank space instead
\else\message{(FIGURES WILL BE INCLUDED)}%
\gdef\figin##1{##1}\gdef\figins##1{##1}\fi}
\def\DefWarn#1{}
\def\figinsert{\goodbreak\midinsert}
\def\ifig#1#2#3{\DefWarn#1\xdef#1{fig.~\the\figno}
\writedef{#1\leftbracket fig.\noexpand~\the\figno}%
\figinsert\figin{\centerline{#3}}\medskip\centerline{\vbox{
\baselineskip12pt\advance\hsize by -1truein
\noindent\footnotefont{\bf Fig.~\the\figno:} #2}}
%\bigskip
\endinsert\global\advance\figno by1}
%%%
\else
\def\ifig#1#2#3{\xdef#1{fig.~\the\figno}
\writedef{#1\leftbracket fig.\noexpand~\the\figno}%
%\figinsert\figin{\centerline{#3}}\medskip
%\centerline{\vbox{\baselineskip12pt
%\advance\hsize by -1truein\noindent
%\footnotefont{\bf Fig.~\the\figno:} #2}}
%\bigskip\endinsert
\global\advance\figno by1} \fi

\def\id{{1 \kern-.28em {\rm l}}}

\def\K3{{\bf K3}}
\def\journal#1&#2(#3){\unskip, \sl #1\ \bf #2 \rm(19#3) }
\def\andjournal#1&#2(#3){\sl #1~\bf #2 \rm (19#3) }

\def\bar{\overline}
\def\hat{\widehat}
\def\ie{{\it i.e.}}
\def\eg{{\it e.g.}}

\def\tilde{\widetilde}

\def\frac#1#2{{#1\over#2}}

\def\d{\partial}

\def\inbar{\,\vrule height1.5ex width.4pt depth0pt}
\def\IC{\relax\hbox{$\inbar\kern-.3em{\rm C}$}}
\def\IR{\relax{\rm I\kern-.18em R}}
\def\IP{\relax{\rm I\kern-.18em P}}

%
%%%%%%%%%%%%%%%%%%%%%%%%%%%%%%%%%%%%
%

%
\catcode`\@=11
\def\slash#1{\mathord{\mathpalette\c@ncel{#1}}}
\overfullrule=0pt

\def\underrel#1\over#2{\mathrel{\mathop{\kern\z@#1}\limits_{#2}}}

\catcode`\@=12

%%%%%%%%%%%%%%%%%%%%%%%%%%%%%%%%%%%%%%%%%%%%%%%%%%%%%%%%%%%%%%

%

\def\tr{{\rm tr}}

\def \sinh{{\rm sinh}}
\def \cosh{{\rm cosh}}

%%%%%%%%%%%%%%%%%%%%%%%%%%%%%%%%%%%%%%%%%%%%%%%%%%%%%%%%%%%%%%
% new defs:

\def\ie{{\it i.e.}}
\def\eg{{\it e.g.}}

%%%%%%%%%%%%%%%%%%%%%%%%%%%%%%%%%%%%%%%%%%%%%%%%%%

%\cite{Asrat:2024jlt}
%\bibitem{Asrat:2024jlt}
\lref\AsratA{
M.~Asrat,
``Rotating strings and anomalous dimensions in Non-AdS holography,''
[arXiv:2404.16601 [hep-th]].
%1 citations counted in INSPIRE as of 03 May 2026
}

%\cite{Asrat:2023yzy}
%\bibitem{Asrat:2023yzy}
\lref\AsratB{
M.~Asrat,
``Moving holographic boundaries,''
Nucl. Phys. B {\bf 1008}, 116699 (2024)
doi:10.1016/j.nuclphysb.2024.116699
[arXiv:2305.15744 [hep-th]].
%5 citations counted in INSPIRE as of 03 May 2026
}

%\cite{Giveon:2017nie}
%\bibitem{Giveon:2017nie}
\lref\Giveon{
A.~Giveon, N.~Itzhaki and D.~Kutasov,
``$T{\bar T} $ and LST,''
JHEP {\bf 07}, 122 (2017)
doi:10.1007/JHEP07(2017)122
[arXiv:1701.05576 [hep-th]].
%178 citations counted in INSPIRE as of 01 May 2023
}

%\cite{Asrat:2024cug}
%\bibitem{Asrat:2024cug}
\lref\AsratC{
M.~Asrat,
``Kalb-Ramond field, black holes and black strings in (2 + 1)D,''
JHEP {\bf 08}, 135 (2025)
doi:10.1007/JHEP08(2025)135
[arXiv:2410.07580 [hep-th]].
%5 citations counted in INSPIRE as of 03 May 2026
}

%\cite{Kutasov:1999xu}
%\bibitem{Kutasov:1999xu}
\lref\KutasovN{
D.~Kutasov and N.~Seiberg,
``More comments on string theory on AdS(3),''
JHEP {\bf 04}, 008 (1999)
doi:10.1088/1126-6708/1999/04/008
[arXiv:hep-th/9903219 [hep-th]].
%330 citations counted in INSPIRE as of 10 May 2026
}

%\cite{DeVega:1992xc}
%\bibitem{DeVega:1992xc}
\lref\DeVega{
H.~J.~De Vega and N.~G.~Sanchez,
``Exact integrability of strings in D-Dimensional De Sitter space-time,''
Phys. Rev. D {\bf 47}, 3394-3405 (1993)
doi:10.1103/PhysRevD.47.3394
%155 citations counted in INSPIRE as of 10 May 2026
}

%\cite{Larsen:1996gn}
%\bibitem{Larsen:1996gn}
\lref\Larsen{
A.~L.~Larsen and N.~G.~Sanchez,
``Sinh-Gordon, cosh-Gordon and Liouville equations for strings and multistrings in constant curvature space-times,''
Phys. Rev. D {\bf 54}, 2801-2807 (1996)
doi:10.1103/PhysRevD.54.2801
[arXiv:hep-th/9603049 [hep-th]].
%52 citations counted in INSPIRE as of 10 May 2026
}

%\cite{Pohlmeyer:1975nb}
%\bibitem{Pohlmeyer:1975nb}
\lref\Pohlmeyer{
K.~Pohlmeyer,
``Integrable Hamiltonian Systems and Interactions Through Quadratic Constraints,''
Commun. Math. Phys. {\bf 46}, 207-221 (1976)
doi:10.1007/BF01609119
%677 citations counted in INSPIRE as of 10 May 2026
}

%\cite{Jevicki:2007aa}
%\bibitem{Jevicki:2007aa}
\lref\Jevicki{
A.~Jevicki, K.~Jin, C.~Kalousios and A.~Volovich,
``Generating AdS String Solutions,''
JHEP {\bf 03}, 032 (2008)
doi:10.1088/1126-6708/2008/03/032
[arXiv:0712.1193 [hep-th]].
%108 citations counted in INSPIRE as of 10 May 2026
}

%\cite{Grigoriev:2007bu}
%\bibitem{Grigoriev:2007bu}
\lref\Grigoriev{
M.~Grigoriev and A.~A.~Tseytlin,
``Pohlmeyer reduction of AdS(5) x S**5 superstring sigma model,''
Nucl. Phys. B {\bf 800}, 450-501 (2008)
doi:10.1016/j.nuclphysb.2008.01.006
[arXiv:0711.0155 [hep-th]].
%148 citations counted in INSPIRE as of 10 May 2026
}

%\cite{Asrat:2024cug}
%\bibitem{Asrat:2024cug}
\lref\MAsrat{
M.~Asrat,
``Kalb-Ramond field, black holes and black strings in (2 + 1)D,''
JHEP {\bf 08}, 135 (2025)
doi:10.1007/JHEP08(2025)135
[arXiv:2410.07580 [hep-th]].
%5 citations counted in INSPIRE as of 10 May 2026
}

%\cite{Maillet:1986vk}
%\bibitem{Maillet:1986vk}
\lref\Maillet{
J.~M.~Maillet,
``New Algebraic Canonical Structures of Integrability in 2-D Field Theories,''
PAR-LPTHE-86/40.
%0 citations counted in INSPIRE as of 11 May 2026
}

%\cite{Babelon:2003qtg}
%\bibitem{Babelon:2003qtg}
\lref\Babelon{
O.~Babelon, D.~Bernard and M.~Talon,
``Introduction to Classical Integrable Systems,''
Cambridge University Press, 2003,
ISBN 978-0-521-03670-2, 978-0-511-53502-4
doi:10.1017/CBO9780511535024
%100 citations counted in INSPIRE as of 11 May 2026
}

\lref\Sagle{
Sagle, Arthur A. “Malcev Algebras.” Transactions of the American Mathematical Society {\bf 101}, no. 3 (1961): 426–58. https://doi.org/10.2307/1993472.
}

%\cite{Semenov-Tian-Shansky:1983fjv}
%\bibitem{Semenov-Tian-Shansky:1983fjv}
\lref\Semenov{
M.~A.~Semenov-Tian-Shansky,
``What is a classical r-matrix?,''
Funct. Anal. Appl. {\bf 17}, 259-272 (1983)
doi:10.1007/BF01076717
%273 citations counted in INSPIRE as of 16 May 2026
}

%\cite{Sklyanin:1982tf}
%\bibitem{Sklyanin:1982tf}
\lref\Sklyanin{
E.~K.~Sklyanin,
``Some algebraic structures connected with the Yang-Baxter equation,''
Funct. Anal. Appl. {\bf 16}, 263-270 (1982)
doi:10.1007/BF01077848
%289 citations counted in INSPIRE as of 17 May 2026
}

%\Bibitem{Mal55}
% A.~I.~Mal'tsev
\lref\Maltsev{
A.~I.~Mal'tsev,
``Analytic loops,"
Mat. Sb. (N.S.) {\bf 36(78)} 3, 569--576 (1955)
}

%\Bibitem{Mal55}
% A.~I.~Mal'tsev
\lref\Kuz{
E.~N.~Kuz'min,
``The connection between Mal'cev algebras and analytic Moufang loops,"
Algebra Logika {\bf 10} 1, 3--22 (1971)
}

\lref\paal{
Paal~Eugen,
``Note on analytic Moufang loops,"
Algebra Logika {\bf 10} 1, 3--22 (1971)
Commentationes Mathematicae Universitatis Carolinae {\bf 45} 2, 349-354 (2004)
}

%\cite{Bianchi:2001pja}
%\bibitem{Bianchi:2001pja}
\lref\Bianchi{
L.~Bianchi,
``On the Three-Dimensional Spaces Which Admit a Continuous Group of Motions,''
Gen. Rel. Grav. {\bf 33}, no.12, 2171-2253 (2001)
doi:10.1023/A:1015357132699
%74 citations counted in INSPIRE as of 19 Aug 2026
}

%\cite{Ellis:1968vb}
%\bibitem{Ellis:1968vb}
\lref\Ellis{
G.~F.~R.~Ellis and M.~A.~H.~MacCallum,
``A Class of homogeneous cosmological models,''
Commun. Math. Phys. {\bf 12}, 108-141 (1969)
doi:10.1007/BF01645908
%634 citations counted in INSPIRE as of 19 Aug 2026
}

\lref\Akivis{
Akivis, M.~A.~ 
``Local algebras of a multidimensional three-web,"
 Sib Math J {\bf 17}, 3-8 (1976)
 doi:10.1007/BF00969285
}

\lref\ivanP{
I.~P.~Shestakov,
``Every Akivis algebra is linear,''
Geometriae Dedicata {\bf 77}, 215-223 (1999)
doi:10.1023/A:1005157524168
}

\lref\Hofmann{
K.~ H.~ Hofmann and K.~ Strambach,
``Lie's fundamental theorems for local analytical loops,"
Pacific J. Math. {\bf 123}, 301-327 (1986)
doi:10.2140/pjm.1986.123.301
}

\lref\Kerdman{
Kerdman, F.~S.~ ,
``Analytic moufang loops in the large,"
Algebra and Logic, {\bf 18}, 325-347 (1979)
doi:10.1007/BF01673501
}

\lref\Izquierdo{
J.~M.~ Perez-Izquierdo and I.~P.~ Shestakov,
``An envelope for Malcev algebras,"
Journal of Algebra {\bf 272}, 379-393 (2004)
doi:10.1016/S0021-8693(03)00389-2
}

%%%%%%%%%%%%%%%%%%%%%%%%%%%%%%%%%%%%%%%%%%%%%%%%%%%
\Title{
} {\vbox{
\centerline{An integrable deformation of the}
%\smallskip
\bigskip
{\vbox{
\centerline{sine(sinh)-Gordon model -- Non-Lie algebra} 
}}
}}

%\Title{
%} {\vbox{
%\centerline{An integrable model governed by}
%\bigskip
%{\vbox{
%\centerline{a non-Lie Malcev algebra} %and non-associative algebra
%}}
%}}

%\centerline{Kalb-Ramond field, black holes and } }}%surfaces
%\smallskip
%\centerline{ black strings in (2 + 1)D} }}
%\bigskip
%\centerline{with $T{\bar T}$} }}

\bigskip
%\bigskip
%\centerline{\it Meseret Asrat\foot{London, UK}}
\centerline{\it Meseret Asrat}
%\bigskip
\smallskip
\centerline{London, UK
%International Center for Theoretical Sciences
}
\centerline{%Tata Institute of Fundamental Research
} \centerline{%Bengaluru, KA 560089, India
}

\smallskip

\vglue .3cm

\bigskip

\let\includefigures=\iftrue
\bigskip
\noindent

In this paper we present an integrable model in two dimensions. It is a deformation of the sine(sinh)-Gordon model. We give its Lax connection. We also obtain its (classical) $r$-matrix. It satisfies the classical Yang-Baxter equation. Thus, the model is a classical integrable field theory in two dimensions. The underlying algebra is a ${\cal Z}$-graded non-Lie alternative Akivis algebra. An alternative Akivis algebra is the tangent space at the identity of an analytic alternative loop as a Lie algebra is the tangent space at the identity of a Lie group. We expect the model to be integrable at the quantum level. We also give a family of classically integrable models which are related to the Poisson-Boltzmann equation in two dimensions.

\bigskip

\Date{04/26}

%%%%%%%%%%%%%%%%%%%%%%%%%%%%%%%%%%%%%%%%%%%%%%%%%%%%%%%%%%%%%%%%
%%%%%%%%%%%%%%%%%%%%%%%%%%%%%%%%%%%%%%%%%%%%%%%%%%%%%%%%%%%%%%%%

\newsec{Introduction}

In this paper we study the integrability of a two-dimensional (2D) field theory first obtained in \AsratA. It describes classical string theory in the ${\cal A}_3$ factor of the string background found in \AsratB. The string background has the product topology ${\cal A}_3 \times S^3 \times {\cal X}_4$, where $S^3$ and ${\cal X}_4$ denote, respectively, a three-sphere and a compact four-dimensional manifold. Examples of ${\cal X}_4$ include the Calabi-Yau manifolds $T^4$ and $K3$. The spacetime ${\cal A}_3$ interpolates between $AdS_3$ in the infrared, \ie, the interior of the bulk, and a (an asymptotically linear dilaton) flat spacetime $\IR\times S^1\times \IR$ in the ultraviolet, \ie, the boundary of the spacetime. The world-sheet conformal field theory (CFT) on ${\cal A}_3$ is an exact marginal current bilinear deformation of the world-sheet theory on $AdS_3$. In the boundary theory the deformation is equivalent to a deformation by an irrelevant operator of (left and right) scaling dimensions $(2, 2)$,\foot{Note irrelevance is defined with respect to a fixed point. An irrelevant operator is less important in the infrared but becomes more important in the ultraviolet.} see \refs{\AsratB,\ \Giveon} and references therein. The resulting deformed theory is thus non-conformal and dual to string theory on ${\cal A}_3$ (times the internal product manifold $S^3 \times {\cal X}_4$).

Although the operator has the same scaling dimensions as the determinant of the energy momentum stress tensor, commonly referred to as $T{\bar T}$, at the (infrared) fixed point, it is not the same operator \KutasovN.\foot{For more and detailed discussion on this see \AsratB.} Moreover, unlike $T{\bar T}$, it is not known how to define it in a generic quantum field theory, nor to what it changes or deforms the theory into. One of the goals of this paper is to shade light on the latter. We achieve this by studying the relation between classical string theory and integrable models.

String theory on Anti-de Sitter (AdS) spacetimes are related to two-dimensional (2D) integrable field theories \refs{\DeVega\Larsen\Jevicki-\Grigoriev}. This is shown primarily by applying Pohlmeyer reduction \Pohlmeyer. For example, classical string theory on $AdS_2$ is equivalent to Liouville theory. Another example which is relevant here is that classical string theory on $AdS_3$ is equivalent to the sinh-Gordon theory. The sinh-Gordon theory is an analytic continuation of the well-known sine-Gordon theory. Other examples include the $B_2$ Toda and a generalized sine-Gordon (non-abelian Toda) theories which are equivalent to classical string theory on $AdS_4$ and $AdS_5\times S^5$, respectively.

Thus, in ${\cal A}_3$ classical string theory is equivalent to a deformed sinh-Gordon theory. A priori, since the spacetime deformation is irrelevant, it is not obvious that the deformed sinh-Gordon theory is also integrable, or most importantly, what the underlying algebraic structure is. In this paper we show explicitly that it is (classically) integrable. We also show that the underlying algebra is a (an infinite dimensional) ${\cal Z}$-graded non-Lie alternative Akivis algebra. Loosely speaking, it is a direct sum of $sl(2, \IR)$ Lie algebras with a shared common Cartan. We will be precise later in the paper. The associated $r$-matrix satisfies the classical Yang-Baxter equation (CYBE). Therefore, smooth functions on the phase space equipped with the Sklyanin bracket form a Lie algebra. Note, this is so though the underlying algebra is not a Lie algebra.

The paper is organized as follows. In section two we review the string background ${\cal A}_3$ first obtained in \AsratB. We also give the deformed sinh-Gordon theory \AsratA\ that describes classical string theory in ${\cal A}_3$. In section three we review the sinh-Gordon theory. In particular we give its Lax connection and (classical) $r$-matrix. As we go through the review, we also discuss integrability more generally in relation to models with non-associative underlying algebras such as a non-Lie Akivis algebra. We in particular discuss the features of the associated Lax equations and Poisson brackets. In section four we discuss the ${\cal Z}$-graded non-Lie alternative Akivis algebra. An alternative Akivis algebra is the tangent space at the identity of an analytic alternative loop as a Lie algebra is the tangent space at the identity of a Lie group.\foot{A loop is a quasigroup with an identity element. It is not necessarily associative. Every group is a quasigroup. Thus, loosely speaking, a loop is a group which is not necessarily associative.} We also construct the Lax connection for the deformed sinh-Gordon theory. We also show that the associated (classical) $r$-matrix satisfies the CYBE. Thus, we in turn show that the theory is (classically) integrable. In section five we discuss more integrable models and future research directions. In Appendix A we show the equivalence between an auxiliary linear problem and the flatness condition. In Appendix B we give the defining relations of the monodromy matrix, or more generally, the state-transition matrix.

\newsec{The string background ${\cal A}_3$}

We now consider the string background ${\cal A}_3$ obtained in \AsratB.\foot{For discussions in relation to black hole and black string solutions in ${\cal A}_3$ see \MAsrat.} ${\cal A}_3$ contains the metric $g_{\mu\nu}$, the Kalb-Ramond field $B_{\mu\nu}$ and the dilaton field $\Phi$. The string metric $g_{\mu\nu}$ is given by
\eqn\aaa{\eqalign{
ds^2 & = g_{ab}dx^adx^b,\cr
& = l^2(d\theta^2 - e^{2\phi}\cosh^2\theta d\varphi^2 + e^{2\phi}\sinh^2\theta d\psi^2),
}
}
where
\eqn\bbb{
e^{-2\phi} = 1 + \gamma^2 - 2\gamma\cosh (2\theta),
}
and $\gamma$ is a dimensionless parameter. It is related to the deformation coupling $\hat\gamma$ in the dual boundary field theory by $\hat\gamma = \gamma R^2$ (up to a constant factor) here $R$ is the conformal radius of the boundary cylinder on which the undeformed CFT is defined.\foot{The conformal boundary of $AdS_3$ (\ie, the universal cover) is a time-like cylinder $\IR \times S^1$. As usual in AdS, we take $R = 1$ in string units.} $l$ is the radius of curvature at $\gamma = 0$.\foot{It is related to the level of the $sl(2, \IR)$ affine Lie algebra $k$ by $l^2 = l_s^2(k - 2)$, where $l_s$ the string length.} The two form $B$ is given by
\eqn\ccc{
B = B_{01}d\varphi\wedge d\psi, \quad B_{01} = -\frac{1}{2}l^2e^{2\phi}(\gamma - \cosh(2\theta)).
}
The dilaton $\Phi$ is given by 
\eqn\ddd{
e^{2\Phi} = g_s^2 |e^{2\phi}| = {g_s^2\over \left|1 + \gamma^2 - 2\gamma \cosh(2\theta)\right|},
}
where $g_s$ is the string coupling at $\gamma = 0$. Note the absolute value sign is necessary in the case $\gamma$ is positive to ensure the dilaton is real everywhere. More on this momentarily but see also \MAsrat. The Kalb-Ramond field strength $H$ is given by
\eqn\dddx{\eqalign{
H & = dB,\cr
& = l^2 (1 - \gamma^2)\sinh(2\theta)e^{4\phi} d\varphi\wedge d\psi\wedge d\theta.
}
}
The time coordinate $\varphi$ takes its value in $\IR$. The radial coordinate $\theta$ takes its value (in the case $\gamma \leq 0$) in $\IR^+$, \ie\ $0\leq \theta < \infty$. The angular variable $\psi$ is periodic. It has period $2\pi(1 - \gamma)$ to ensure that there is no conical singularity at $\theta = 0$. The deformation coupling $\gamma$ takes the values in the range
\eqn\zf{-1 \leq \gamma \leq 1.
}
Other values of $\gamma$ can be mapped onto the interval \zf\ by rescaling the coordinates and the string coupling at $\gamma = 0$, \ie, $g_s$. The transformation rules under $\gamma \to 1/\gamma$ are $\varphi \to \varphi/\gamma$, $\psi \to -\psi/\gamma$ and $g_s^2 \to g_s^2/\gamma^2$ (for more details see \refs{\AsratB, \MAsrat}).\foot{Thus, the transformation $\gamma \to {1\over\gamma}$ is a strong-to-weak coupling mapping. $g_s$ becomes smaller.} The metric, the three form antisymmetric tensor and dilaton are invariant by the transformations. 

We note that the dilaton is related to the Poisson kernel for a unit disk by an analytic continuation. The Poisson kernel is given by ${\cal P}_\gamma(i\theta)$ where ${\cal P}_\gamma$ is proportional to $e^{2\phi}$ and it is given by
\eqn\vvv{{\cal P}_\gamma(s) = {1 - \gamma^2 \over 1 + \gamma^2 - 2\gamma \cosh(s)}, \quad {\cal P}_{1\over \gamma}(s) = -{\cal P}_{\gamma}(s), \quad {\cal P}_\gamma(s) = {\cal P}_\gamma(-s).
} 
In the kernel ${\cal P}_\gamma(i\theta)$, $\theta$ has period $2\pi$. $\gamma$ is in the interval $[0,1)$ since  ${\cal P}_\gamma(i\theta) = {\cal P}_{-\gamma}\left(i(\theta +\pi)\right)$.

In this paper we mostly consider the case where $\gamma$ is negative. It corresponds to washing away the conformal boundary of $AdS_3$ to infinity. In the case $\gamma$ is positive the deformation creates at onset a curvature singularity at the location of the conformal boundary. Therefore, in holographic theories with time-like boundary, the deformation with positive $\gamma$ is not defined \AsratB. It implies that we cannot localize gravity in a finite size box. We comment more on this later in the paper.

As we noted earlier, with several examples, classical string solutions on AdS are mapped to the generalized sinh-Godron theory via the Pohlmeyer reduction. The Pohlmeyer reduction is a non-local and many-to-one transformation. Therefore, it is enough to consider a particular classical string solution to identify the integrable theory. Since ${\cal A}_3$ is a deformation of $AdS_3$ and also the reduction does not apply directly because the deformation is irrelevant in spacetime, we consider a rotating string solution to identify the deformation on the sinh-Gordon theory.

The deformed equation of motion is \AsratA,
\eqn\aaaxx{\partial^2 \theta - {2\over 1-\gamma^2} \cdot \partial_\theta {\cal P}_\gamma(\theta) = 0, \quad \gamma \neq 0.
}
At $\gamma = 0$ or for small $\gamma$, \ie, $|\gamma| \ll 1$, it is the sinh-Gordon equation of motion \AsratA. In this paper we study a more general relativistic two-dimensional (2D) model with the equation of motion given by
\eqn\aaax{\partial^2 \theta - \alpha^{-2} \partial_\theta {\cal P}_\gamma(2\beta\theta) = 0,
}
where $\alpha$ is a smooth function of only $\gamma$, and $\beta$ is a positive constant independent of $\gamma$. The analytic continuation $\theta \to i\theta$ in \aaax\ gives the corresponding equation of motion of the deformed sine-Gordan model. It is given by
\eqn\aaaz{\partial^2 \theta + \alpha^{-2} \partial_\theta {\cal P}_\gamma(2\beta i \theta) = 0.
}

We next discuss the sinh-Gordon model. We give its Lax connection and $r$-matrix. The discussion is more general and it applies to models with non-associative underlying algebras. We use the results as a starting point to obtain the Lax connection and $r$-matrix for the deformed model.

\newsec{The sinh-Gordon model}

The 2D sinh-Gordon model is described by the Lagrangian density
\eqn\aaa{{\cal L} = {1\over 2}\partial_\alpha\phi \partial^\alpha \phi - {m^2\over \beta^2}\left[\cosh(\beta\phi) - 1\right].
}
The equation of motion is given by
\eqn\bbb{\partial^2\phi + {m^2\over \beta}\sinh(\beta\phi) = 0.
}

We introduce the light cone coordinates
\eqn\ccc{u = {1\over 2}\left(x + t\right), \quad v = {1\over 2}\left(x - t\right).
}
We use the metric 
\eqn\ddd{ds^2 = dt^2 - dx^2 = -4dudv.
}
Thus, we have the relations
\eqn\eee{\partial_t = {1\over 2}\left(\partial_u - \partial_v\right), \quad \partial_x = {1\over 2}\left(\partial_u + \partial_v\right), \quad \partial^2 = \partial^2_t - \partial^2_x = -\partial_u\partial_v.
}
In terms of the light cone coordinates the equation of motion takes the form
\eqn\fff{\phi_{uv} - {m^2\over \beta}\sinh(\beta\phi) = 0.
}

Two-dimensional (2D) integrable field theories are characterized by a Lax connection or pair $(A_u, A_v)$ with values in some algebra.\foot{In this paper it will be primarily the Akivis algebra. Non-Lie Akivis algebra does not satisfy the Jacobi identity, and thus, it is not representable as a matrix algebra. Because matrix multiplication is associative, a matrix algebra (in which the matrix multiplication is replaced by a commutator) automatically satisfies the Jacobi identity. (The action on the algebra itself can be however represented by matrices. This is the adjoint representation. In general it is not a Lie homomorphism.)} $A_u$ and $A_v$ will in general depend on a free parameter $\lambda$ known as the spectral parameter. On-shell, for all $\lambda$, the pair solves the flatness or zero-curvature condition,
\eqn\ggg{\partial_u A_v - \partial_v A_u + \left[A_u, A_v\right] = 0.
}
Thus, the zero-curvature condition is equivalent to the field equations.\foot{The curvature is the two-form $F = dA + A \wedge A$. In components  $F = {1\over 2}F_{\mu\nu} dx^\mu \wedge dx^\nu, F_{\mu\nu} = \partial_\mu A_\nu - \partial_\nu A_\mu + \left[A_\mu, A_\nu\right]$.} It also implies that locally the path-ordered exponential of the connection (integrated along a path) depends only on the endpoints of the path. We are assuming (the existence of) a correspondence similar to that between a Lie algebra and a Lie group. More on this momentarily. Note that the commutator in \ggg\ is taken in the underlying algebra. In the case the algebra is a Lie algebra, $A_u$ and $A_v$ are in general represented by matrices which form the defining or fundamental representation.

The flatness condition follows from a compatibility condition of an auxiliary problem. The auxiliary problem is
\eqn\o{\partial_u\psi = A_u\psi, \quad \partial_v\psi = A_v\psi.
}
It is linear in $(u, v)$. $\psi$ is called the wave function. $A_u$ and $A_v$ can be thought of as the $u$ and $v$ components of the Lax connection. The compatibility condition is $ (\partial_u\partial_v)\psi= \partial_u(\partial_v\psi) = \partial_v(\partial_u\psi)$.

In the spacetime coordinates, the flatness condition is equivalent to
\eqn\cc{\partial_t L - \partial_x M + \left[L, M\right] = 0, 
}
where
\eqn\ccc{L = {1\over 2}\left(A_u + A_v\right), \quad M = {1\over 2}\left(A_u - A_v\right).
}
The corresponding auxiliary problem linear in $(x, t)$ is (see Appendix A)
\eqn\oozxzoo{\partial_x\psi = L\psi, \quad \partial_t\psi = M\psi.
}

The wave function $\psi$ belongs to the module (or carrier space) over the (universal enveloping) algebra, and it gives, at fixed $t$, \ie, on a spatial slice, the monodromy matrix $T(\lambda, t)$ (or more generally the state-transition matrix $T(x, y, \lambda)$)\foot{The monodromy matrix, which is defined in periodic models or models living on an infinite cylinder $\IR \times S^1$, where $S^1$ denotes the spatial circle, is the state-transition matrix after one cycle, see Appendix B.} which generates the infinite set of conserved quantities or charges \refs{\Babelon, \Maillet}.

Unless the underlying algebra is (an associative algebra or) a Lie algebra, in general $T$ is not a matrix since matrix multiplication is associative. However, we will still refer to it as a monodromy matrix (or more generally as a state-transition matrix). In the case the spatial direction is a circle, it is regarded as a holonomy at the identity. In this paper, it belongs to an analytic alternative loop. The tangent space at the identity of an analytic alternative loop is an alternative Akivis algebra.  We denote the identity by $e$.

The monodromy matrix (or more generally the state-transition matrix) $T$ is given by the path-ordered exponential of the Lax connection (spatial) component $L$. In non-associative algebra (or analytic loop), to define the order of operations, we choose a particular nesting prescription of parentheses. For example, in a right-nested prescription, $w x y z := w\cdot(x\cdot(y\cdot z))$, where the dot $(\cdot)$ represents the algebra multiplication operation.

The monodromy matrix $T$ satisfies (with periodic boundary conditions) the Lax equation
\eqn\ttm{{\partial T\over \partial t} = \rho_T M - \lambda_T M,
}
where $\rho_T := (dR_T)_e$ and $\lambda_T:= (dL_T)_e$ denote differentials of the right and left translations by $T$, respectively \refs{\Semenov}. $T \in $ an analytic alternative loop \refs{\Hofmann}. The tangent spaces at different points on the analytic loop are isomorphic vector spaces. We in particular note that $\partial T/\partial t$ lives in the tangent space at $T$. Therefore, it can be written as a commutator. The Lax equation \ttm\ is one of the relations that define $T$.\foot{More generally, one can take \ttm\ as the (defining) definition of the monodromy matrix.} Appendix B contains more relations which together (alternatively) define $T$.

Under periodic boundary conditions, the conserved charges are essentially obtained by applying the trace (invariant form) on the algebra. In general given a (an endomorphism) linear transformation, we can define a trace. A (well-defined) trace has the property that it gives zero on any commutator. It follows in particular that the algebras we consider consist of zero trace elements.\foot{This is shown more generally using the identity $\tr([z,[x, y]]) = \tr([[y,z],x])$ for elements $x, y, z$ of the algebras \Sagle.} Thus, $\tr(\partial T/\partial t) = 0$. Since the flatness condition holds for all values of the spectral parameter $\lambda$, expanding the trace $\tr(T)$ in $\lambda$ gives an infinite set of conserved charges. When the spatial coordinate is not on a circle but on a line, it is the infinite volume limit of the state-transition matrix $T(x, y, \lambda)$, \ie, $T := T(-\infty, \infty, \lambda)$, with proper regularization, that gives the conserved charges.

For the sinh-Gordon theory the connection components are given by\foot{Note Lax pairs are not unique.}
\eqn\hhh{\eqalign{
A_u & = {\beta\over 4}\phi_u \sigma_3 + {m\lambda\over 2}\left(e^{\beta\phi/2} \sigma_+ - e^{-\beta\phi/2}\sigma_-\right),\cr
A_v & = -{\beta\over 4}\phi_v \sigma_3  + {m\over 2 \lambda}\left(e^{-\beta\phi/2} \sigma_+ - e^{\beta\phi/2}\sigma_-\right),
}
}
where $\sigma_3$, $\sigma_+$ and $\sigma_-$ obey the $sl(2, \IR)$ commutation relations
\eqn\iiiaaa{\left[\sigma_3, \sigma_\pm\right] = \pm 2\sigma_\pm, \quad \left[\sigma_-, \sigma_+\right] = \sigma_3.
}
Thus, the underlying algebra is the $sl(2, \IR)$ Lie algebra.\foot{We are taking $\lambda$ to be real (but it can also be complex). The fundamental representation is given by a linear combination of the Pauli spin matrices,
\eqn\iii{\sigma_3 = \pmatrix{ 1 & 0\cr
0 & -1}, \quad \sigma_+ = \pmatrix{ 0 & 1\cr
0 & 0}, \quad \sigma_- = \pmatrix{ 0 & 0\cr
-1 & 0}.
}
}
We note that
\eqn\jjj{\eqalign{\partial_u A_v & = -{\beta\over 4}\phi_{uv} \sigma_3  + {m\beta\over 4 \lambda}\left(-e^{-\beta\phi/2} \sigma_+ - e^{\beta\phi/2}\sigma_-\right)\phi_u,\cr
\partial_v A_u & = {\beta\over 4}\phi_{uv} \sigma_3  + {m\beta\lambda\over 4 }\left(e^{\beta\phi/2} \sigma_+ + e^{-\beta\phi/2}\sigma_-\right)\phi_v,\cr
\left[A_u, A_v\right] & =  {m^2\over 2} \sinh(\beta\phi)\sigma_3 + {m\beta\over 4 \lambda}\left(e^{-\beta\phi/2}\sigma_+ + e^{\beta\phi/2}\sigma_-\right)\phi_u + {m\beta\lambda\over 4}\left(e^{\beta\phi/2}\sigma_+ + e^{-\beta\phi/2}\sigma_-\right)\phi_v,\cr
\partial_u A_v - \partial_v A_u & = -{\beta\over 2}\phi_{uv} \sigma_3  - {m\beta\over 4 \lambda}\left(e^{-\beta\phi/2} \sigma_+ + e^{\beta\phi/2}\sigma_-\right)\phi_u - {m\beta\lambda\over 4 }\left(e^{\beta\phi/2} \sigma_+ + e^{-\beta\phi/2}\sigma_-\right)\phi_v.
}
}
Therefore, the zero-curvature condition gives for all $\lambda$,
\eqn\kkk{-{\beta\over 2}\phi_{uv} + {m^2\over 2}\sinh(\beta\phi) = 0,
}
which is the equation of motion \fff. This ensures that the monodromy $T(t, \lambda)$ exists and that there are infinite set of conserved charges. However, integrability requires also the charges to be in involution. A sufficient condition, in an arbitrary algebra, is that the Poisson bracket of $L$ takes certain forms or structures.\foot{A bracket is a Poisson bracket only if it also satisfies, among other properties, the Jacobi identity.} The two most common structures are the Sklyanin and Maillet structures or brackets \refs{\Maillet, \Sklyanin}.

For the sinh-Gordon Model the spatial or $x$ component of the Lax connection is 
\eqn\dd{L = L^a_\lambda\sigma_a = {\beta\over 4}\pi \sigma_3 +  {m\over 4}\left(\alpha^+(\phi)\sigma_+ - \alpha^-(\phi)\sigma_-\right),
}
where
\eqn\ee{\pi = \partial_t \phi, \quad \alpha^\pm = \lambda e^{\pm \beta\phi/2} + {1\over \lambda} e^{\mp \beta\phi/2}.
}
We denote 
\eqn\ff{L_1 = L \otimes I, \quad L_2 = I\otimes L.
}
The index 1 or 2 means that $L$ sits in the first or second factor in the tensor product of two $sl(2,\IR)$.\foot{The tensor notation is for convenience, and it applies more generally to any algebra.} The Poisson bracket is
\eqn\gg{\left\{L_1, L_2\right\} = \left\{L^a_\lambda(x), L^b_\mu(y)\right\}\left(\sigma_a\otimes \sigma_b\right).
}
We have the defining canonical equal-time relations
\eqn\hh{\left\{\phi(x), \pi(y)\right\} = \delta(x - y), \quad \left\{\phi(x), \phi(y)\right\} = 0, \quad \left\{\pi(x), \pi(y)\right\} = 0.
}
Using \hh\ we get
\eqn\kk{\eqalign{\left\{L_1, L_2\right\} & = \left[{\tilde\alpha^+}(\lambda)(\sigma_+\otimes \sigma_3) - {\tilde\alpha^+}(\mu)(\sigma_3\otimes \sigma_+) + {\tilde\alpha^-}(\lambda)(\sigma_-\otimes \sigma_3) - {\tilde\alpha^-}(\mu)(\sigma_3\otimes \sigma_-)\right] \delta(x-y),
}
}
where
\eqn\kkx{{\tilde\alpha}^\pm(\nu) = {m\beta^2\over 32}\left(\nu e^{\pm \beta \phi(x)/2} - {1\over \nu}e^{\mp \beta \phi(x)/2}\right).
}
Thus, the bracket has the ultralocal or Sklyanin structure. It is ultralocal due to the solely presence of the distribution $\delta(x-y)$. It does not contain derivatives of the distribution. It spans a large class of integrable field theories \Babelon.

To determine the $r$-matrix, we look for a bracket of the form
\eqn\llzxzll{\left\{L_1, L_2\right\} = \left[r_{12}(\lambda, \mu), L_1(\lambda) + L_2(\mu)\right]\delta(x-y),
}
$r_{12}$ acts on the tensor product $V_1\otimes V_2$. $V_1$ and $V_2$ are copies of $sl(2,\IR)$. We write
\eqn\mm{r_{12} = r_{12}^{ab}\left(\sigma_a \otimes \sigma_b\right).
}
Therefore, we have
\eqn\nn{\eqalign{\left[r_{12}, L_1\right] & = r^{ab}_{12}L_\lambda^{c}\left[\sigma_a, \sigma_c\right]\otimes \sigma_b,\cr
\left[r_{12}, L_2\right] & = \sigma_a \otimes \left[\sigma_b, \sigma_c\right]r^{ab}_{12}L_\mu^{c}. 
}
}
We find
\eqn\qq{\eqalign{r^{33}_{12} & = {\beta^2\over 16}{\lambda^2 +\mu^2\over \lambda^2 -\mu^2},\cr
r^{+-}_{12} & = r^{-+}_{12} = -{\beta^2\over 4}{\lambda \mu\over \lambda^2-\mu^2}.
}
}
All the other components are zero. Thus, the $r$-matrix is 
\eqn\rr{r_{12}(\lambda, \mu) = {\beta^2\over 4}\left\{{1\over 4}{\lambda^2 +\mu^2\over \lambda^2 -\mu^2}(\sigma_3 \otimes \sigma_3) - {\lambda \mu\over \lambda^2-\mu^2}(\sigma_- \otimes \sigma_+ + \sigma_+ \otimes \sigma_-)\right\} = -r_{21}(\mu, \lambda).
}
It is non-dynamical, \ie, it is independent of the dynamical fields and spacetime coordinates. For \kk\ to define a Poisson bracket it should satisfy Jacobi identity. In order for the Poisson brackets to satisfy the Jacobi identity it is sufficient that the (constant) $r$-matrix \rr\ satisfy the CYBE. Note this is independent of the underlying algebra. For example, the algebra may not satisfy the Jacobi identity.

The classical Yang-Baxter equation (CYBE) is
\eqn\ss{[r_{12}, r_{13}] + [r_{12}, r_{23}] + [r_{13}, r_{23}] = 0,
}
where $r_{ij}$ stands for $r_{ij}(\lambda_i,\lambda_j)$ and
\eqn\tt{
r_{12} = r^{ab}_{12}\left(\sigma_a\otimes \sigma_b \otimes I\right), \quad r_{13} = r^{ab}_{13}\left(\sigma_a\otimes I \otimes \sigma_b \right), \quad r_{23} = r^{ab}_{23}\left(I \otimes \sigma_a\otimes \sigma_b\right).
}
The indices on $r$ refer to the copies of $sl(2,\IR)$ on which $r$ is acting. We note that
\eqn\vv{ \eqalign{2r^{33}_{12}r^{+-}_{13} +  r^{+-}_{12}r^{+-}_{23} + 2r^{+-}_{13}r^{33}_{23} & = 0,\cr
2r^{+-}_{12}r^{33}_{13} - 2r^{+-}_{12}r^{33}_{23} - r^{+-}_{13}r^{-+}_{23} & = 0,\cr
 r^{+-}_{12}r^{-+}_{13} + 2r^{33}_{12}r^{-+}_{23} - 2r^{33}_{13}r^{-+}_{23} & = 0,\cr
}
}
and $r^{+-}_{ij} = r^{-+}_{ij}$. Thus, $[r_{12}, r_{13}] + [r_{12}, r_{23}] + [r_{13}, r_{23}] = 0$. The Poisson bracket \llzxzll\ with an $r$-matrix which satisfies the CYBE is called Sklyanin bracket \Sklyanin. For theories with non-ultralocal properties, Jacobi identity requires a modified Yang-Baxter equation to be satisfied \Maillet. The Sklyanin bracket for the monodromy matrix $T$ is (see Appendix B)
\eqn\zzqqx{
\{T_1(\lambda), T_2(\mu)\} = \rho_{T_1}\rho_{T_2}r_{12}(\lambda, \mu) - \lambda_{T_1}\lambda_{T_2}r_{12}(\lambda, \mu),
}
where $\rho_{T_1} = \rho_T\otimes I, \rho_{T_2} = I \otimes \rho_T, \lambda_{T_1} = \lambda_T\otimes I, \lambda_{T_2} = I \otimes \lambda_T$. 

\newsec{An integrable deformation and a class of non-Lie Akivis algebras}

We now consider the deformed sinh-Gordon model. We first consider an infinite dimensional deformation (or a non-Lie extension) of the $sl(2, \IR)$ Lie algebra.

\subsec{A class of non-Lie Akivis algebras}

Consider an algebra with basis $e_n, n = 0, \pm 1, \pm 2, \cdots$. They obey the commutation relations
\eqn\abb{\eqalign{\left[e_n, e_m\right] & = m\delta_{n + m, 0}e_0, \quad n, m \neq 0,\cr
\left[e_0, e_n\right] & = ne_n, \quad \left[e_n, e_0\right] = -ne_n. 
}
}
The Cartan element $e_0$ induces a grading of the algebra by the adjoint action. We assign a degree to each generator as follows.
\eqn\acc{{\rm deg}(e_n) = n.
}
The zero element is considered a homogeneous element of every degree. Thus, the algebra is a ${\cal Z}$-graded algebra 
\eqn\add{A = \bigoplus_{n \in Z}A_n, \quad e_n \in A_n.
}
We next show that the algebra $A$ is an alternative Akivis algebra. Moreover, we show that it satisfies additional identities. Therefore, it is a particular class of an alternative Akivis algebra.

An Akivis algebra is a non-associative algebra equipped with a bilinear anti-commutative multiplication, the commutator $[x, y]$, and a trilinear ternary multiplication, the associator $[x, y, z]$, that satisfy the Akivis identity \Akivis
\eqn\akivisa{[[x, y], z] + [[y, z], x] + [[z, x], y] = [x, y, z] + [y, z, x] + [z, x, y] - [x, z, y] - [y, x, z] - [z, y, x].
}
In the case the associator is identically zero, the Akivis identity reduces to the Jacobi identity, and thus, the corresponding algebra is a Lie algebra.

Every Akivis algebra admits an isomorphic embedding into a not necessarily associative algebra equipped with a bilinear multiplication $x\cdot y$, where the commutator and associator are defined by $[x, y] := x\cdot y - y\cdot x$ and $[x, y, z] := (x\cdot y)\cdot z - x\cdot (y\cdot z)$, respectively \refs{\ivanP, \Izquierdo}. The algebra is an Akivis algebra with respect to the commutator and associator operations. This is one of the theorems relevant for our discussion.

An Akivis algebra is alternative if and only if the associator is antisymmetric. It follows that for an alternative Akivis algebra we have
\eqn\newaeeee{[x, y, z] = {1\over 6} J(x, y, z),
}
where $J$ is the Jacobian. Thus, the associator is completely determined by the commutator. A standard example of an alternative Akivis algebra is a Malcev algebra (or a Moufang-Lie algebra) \Maltsev. A Malcev algebra additionally satisfies the Malcev identity \Sagle
\eqn\aee{
[[x, y],[x, z]] = [[[x, y], z], x] + [[[y, z], x], x] + [[[z, x], x], y].
}
Note it is not linear in $x$. Every Lie algebra is a Malcev algebra, but not the converse.

Every Lie algebra determines a unique local Lie group. Similarly, every alternative Akivis algebra determines a unique local analytic alternative loop \Hofmann. The loop is unique up to local isomorphism. This is the other theorem relevant for our discussion. A Malcev algebra, for example, corresponds to an analytic Moufang loop \refs{\Maltsev\Kuz\paal-\Kerdman}.

We note that the algebra \abb\ with the associator \newaeeee\ is an alternative Akivis algebra. It does not (in general) satisfy the Jacobi identity.\foot{Take, for example, $(x,y,z) =(e_{-n}, e_{n}, e_{n + 1})$ for any $n \geq 1$. For these elements the Jacobi identity is not satisfied. We also note that $[[x, y],z] -[x,[y,z]] \neq 0$.} Thus, it is a ${\cal Z}$-graded non-Lie alternative Akivis algebra. The algebra additionally satisfies two identities. The first identity is for any elements $x, y$ of the algebra $A$. It is given by 
\eqn\newaeeb{J(x, y, e_0) = 0.
}
The second identity is for any four elements $w, x, y, z$ of the algebra $A$. It is given by
\eqn\newaee{[[J(w, x, y), [z, e_0]], e_0] = 0.
}
We expect a stronger version of the identity \newaee\ to exist. The identities impose constraints on the analytic alternative loop.

The identity \newaee\ is in general different from the Malcev identity \aee. However, the basis $\{e_n, n = 0, \pm 1, \pm 2, \cdots\}$ satisfies the Malcev identity \aee. To show that the  elements $e_n, n = 0, \pm 1, \pm 2, \cdots$ satisfy \aee\ it is enough to consider only the case where $(x, y, z) = (e_l, e_m , e_n)$ with $l,m,n \neq 0$. All the other combinations satisfy the Jacobi identity \newaeeb\ and, in turn, satisfy the Malcev identity. We have
\eqn\opch{\eqalign{[[x, y],[x, z]] & = [[[y, z], x], x] = 0,\cr
 [[[x, y], z], x] & = m\delta_{l+m,0}\cdot n \cdot l \delta_{n + l,0}e_0 = mnl\delta_{l+m,0}\delta_{n + l,0}e_0,\cr
 [[[z, x], x], y] & = l\delta_{n + l, 0}\cdot l\cdot m\delta_{l + m,0}e_0 = ml^2\delta_{l+m,0}\delta_{n + l,0}e_0,\cr
  [[[x, y], z], x]  + [[[z, x], x], y] & = ml(n + l)\delta_{l+m,0}\delta_{n + l,0}e_0 = 0.
}
}
Thus, the Malcev idenity is satisfied.

Although the algebra is a Malcev-like algebra (for the reason that the basis satisfies the Malcev identity), it is not a Malcev algebra. Take, for example, $(x, y, z) = (e_{-2} + e_2, e_1, e_{-1})$.

We note that for any given positive $n \neq 0$, the subset consisting of $A_{-n}, A_0, A_n$ is a Lie algebra \newaeeb. The non-vanishing commutation relations are
\eqn\aoo{[e_n, e_{-n}] = -ne_0, \quad [e_0, e_n] = ne_n, \quad [e_0, e_{-n}] = -ne_{-n}.
}
It is isomorphic to the $sl(2, \IR)$ Lie algebra. Thus, loosely speaking, the algebra $A$ is a direct sum of $sl(2, \IR)$ Lie algebras with a shared common Cartan. The Cartan is $A_0$.\foot{We denote by $A_{m,n}$ the subset consisting of $A_m, A_0, A_n$. We note that for any non-zero $m, n$ with $m \neq n, m + n \neq 0$, the subset $A_{m, n}$ is also a Lie algebra \newaeeb. Its basis elements satisfy the commutation relations
\eqn\aoobin{[e_m, e_n] = 0, \quad [e_0, e_n] = ne_n = -[e_n, e_0], \quad [e_0, e_m] = me_m = -[e_m, e_0].
}
In the standard Bianchi classification (of all the three-dimensional Lie algebras over the field of the real numbers), it belongs to the Bianchi type VI$_h$ family \refs{\Bianchi, \Ellis}. The Lie algebra $A_{n, n}$ is defined shortly below.

The pair consisting of the Lie algebras $A_{m,n}$ and $A_{-m, k}$ with $n \neq k$ forms a non-Lie alternative Akivis algebra. Note, loosely speaking, it is a direct sum of the Lie algebras $A_{n,k}$ and $A_{-m,m}$ with a common $e_0$. $A_{-j,j}$ is the $sl(2, \IR)$ Lie algebra. Thus, loosely speaking, the direct sum of $N \ge 2$ $sl(2, \IR)$ Lie algebras, \ie, $A_{-1,1}, A_{-2,2}, \cdots, A_{-(N-1),N - 1}, A_{-N,N}$, with a common $e_0$ is equivalent, loosely speaking, to the direct sum of the $N$ type VI$_h$ Lie algebras $A_{-1,2}, A_{-2,3}, \cdots, A_{-n,n+1}, A_{-(n+1),n+2}, \cdots, A_{-(N - 1),N}, A_{-N,1}$ with a common $e_0$. Note the equivalence is not unique. For example, for $N = 2$, we have $A_{-1, 2}, A_{-2,1}$ and $A_{1, 2}, A_{-1, -2}$.

Moreover, the subset consisting of $A_n$ and $A_0$ is also a Lie algebra \newaeeb. We denote the subset by $A_{n, n}$. It is the Lie algebra of the affine group of the real line. It is denoted by ${\rm aff}(1, \IR)$. The commutation relations are
\eqn\aaffine{[e_0, e_n] = ne_n = -[e_n, e_0], \quad [e_0, e_0] = 0, \quad [e_n, e_n] = 0.
}
The pair $A_{m, n}$ and $A_{-m, n}$ also forms a non-Lie alternative Akivis algebra. Loosely speaking, it is the direct sum of the Lie algebras $sl(2, \IR)$ and ${\rm aff}(1, \IR)$ with a shared common Cartan.
}

For convenience, we denote the basis of the algebra as $e_n = \sigma^+_n, e_{-m} = \sigma^-_m, e_0 = \sigma_0$ and $m, n \geq 1$. The commutation relations now become
\eqn\ppp{
\left[\sigma_n^{+}, \sigma_m^{-}\right] = -n \delta_{n,m}\sigma_0, \quad \left[\sigma_0, \sigma^{\pm}_n\right] = \pm n \sigma^{\pm}_n, \quad \left[\sigma^{\pm}_n, \sigma^{\pm}_m\right] = 0, \quad \left[\sigma_0, \sigma_0\right] = 0.
}
We note that it is neither a central extension nor an extension by derivation.\foot{Moreover, an extension of a Lie algebra always gives a Lie algebra with dimension bigger than or similar to the original Lie algebra.} See Fig. 1 for a pictorial representation of the structure of the algebra $A$.\foot{In the figure, any simple polygonal chain consists of two (connected) line segments (or edges) represents a Bianchi type VI$_h$ Lie algebra \aoobin\ (see footnote 18). An example is the chain consisting of the elements $\sigma^-_4, \sigma_0, \sigma^+_1$. The elements form an angle and satisfy the commutation relations $[\sigma_0, \sigma^+_1] = +\sigma^+_1$, $[\sigma^-_4, \sigma_0] = +4\sigma^-_4$, and $[\sigma^-_4, \sigma^+_1] = 0$. They generate the subset $A_{-4,1}$. Another example is the path that passes through $\sigma^-_3, \sigma^-_2, \sigma_0$. It represents the subset $A_{-3,-2}$. Each line segment or edge represents the two-dimensional non-abelian Lie algebra ${\rm aff}(1, \IR)$ \aaffine.}
 
 \ifig\loc{The plot depicts the structure of the Malcev algebra $A$. For simplicity, without loss of generality, only a subset of the algebra is shown. The star at the center represents the Cartan, and $\sigma_0$ is the Cartan generator. Along a given line there are only three basis elements. The subscripts on $\sigma^{\pm}_n$ denote the line number. For example, along line $\#4$ we have $\sigma^-_4, \sigma_0, \sigma^+_4$, and along line $\#2$ we have $\sigma^-_2, \sigma_0, \sigma^+_2$. The lines do not intersect except at the center. Thus, $\sigma_0$ is the only element which is shared commonly. Each line represents an $sl(2, \IR)$ Lie algebra. Since $[\sigma_0, \sigma_n^\pm] = \pm n\sigma^\pm_n$ and $[\sigma^-_n, \sigma^+_n] = +n\sigma_0$, the distance from the center represents $n$, \ie, the line number. It is also the magnitude of the degrees of the elements $\sigma^\pm_n$. The direction of the arrow represents the sign of the degree. Away from the center the sign is $+$ and towards the center it is $-$. It also represents the signs of the commutation relations. Only those elements residing on a (straight) line have non-trivial commutation relations. For example, on line $\#1$, along the arrow direction, we have $[\sigma_0, \sigma^+_1] = +\sigma^+_1$, $[\sigma^-_1, \sigma_0] = +\sigma^-_1$, and $[\sigma^-_1, \sigma^+_1] = +\sigma_0$. Any two or more lines form a non-Lie alternative Akivis subalgebra. For a total of $N$ lines it is a ${\cal Z}_{2N + 1}$-graded subalgebra. In the figure $N = 4$. One may refer to $A$ as a hub-and-spoke deformation (or extension) of $sl(2, \IR)$. $\sigma_0$ is the central hub, and all the other $\sigma^\pm_n$ are the spokes.}
{\epsfxsize4.0in\epsfbox{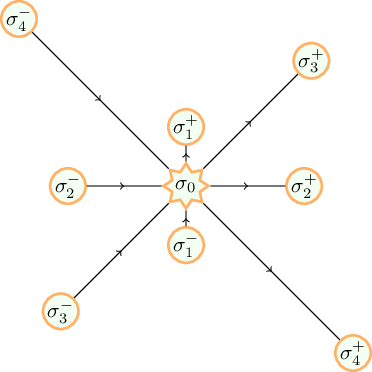}}
 
 \subsec{An integrable model governed by a non-Lie Akivis algebra}
 
Consider the deformed Lax connection components $A_u$ and $A_v$ given by
\eqn\yyy{\eqalign{
A_u & = \beta\phi_u \sigma_0 + {2\lambda\beta\over \alpha}\sum_{n = 1}^{\infty}\gamma^{n/2}\left(e^{n\beta\phi} \sigma^+_n + e^{-n\beta\phi}\sigma^-_n\right),\cr
A_v & = -\beta\phi_v \sigma_0  + {2\beta\over  \lambda \alpha}\sum_{k = 1}^{\infty}\gamma^{k/2}\left(e^{-k\beta\phi} \sigma^+_k + e^{k\beta\phi}\sigma^-_k\right).
}
}
We note that
\eqn\sss{\eqalign{\partial_u A_v & = -\beta\phi_{uv} \sigma_0  + {2\beta^2\over  \lambda \alpha}\sum_{k = 1}^{\infty}k\gamma^{k/2}\left(-e^{-k\beta\phi} \sigma^+_k + e^{k\beta\phi}\sigma^-_k\right)\phi_u,\cr
\partial_v A_u & = \beta\phi_{uv} \sigma_0  + {2\lambda\beta^2\over \alpha }\sum_{n = 1}^{\infty}n\gamma^{n/2}\left(e^{n\beta\phi} \sigma^+_n - e^{-n\beta\phi}\sigma^-_n\right)\phi_v,\cr
\left[A_u, A_v\right] & = -{8\beta^2 \over \alpha^{2}}\sum_{n = 1}^{\infty} n \gamma^{n} \sinh(2n\beta\phi)\sigma_0 +  {2\beta^2\over \lambda\alpha}\sum_{k = 1}^{\infty}k\gamma^{k/2}\left(e^{-k\beta\phi} \sigma^+_k - e^{k \beta\phi} \sigma^-_k \right)\phi_u\cr
& + {2\lambda\beta^2\over \alpha}\sum_{n = 1}^{\infty}n \gamma^{n/2}\left(e^{n\beta\phi}\sigma^+_n - e^{-n\beta\phi}\sigma^-_n\right)\phi_v,\cr
\partial_u A_v - \partial_v A_u & = -2\beta\phi_{uv} \sigma_0  - {2\beta^2\over  \lambda\alpha}\sum_{k = 1}^{\infty}k\gamma^{k/2}\left(e^{-k\beta\phi} \sigma^+_k - e^{k\beta\phi}\sigma^-_k\right)\phi_u \cr
& - {2\lambda\beta^2\over \alpha }\sum_{n = 1}^{\infty}n\gamma^{n/2}\left(e^{n\beta\phi} \sigma^+_n - e^{-n\beta\phi}\sigma^-_n\right)\phi_v.
}
}
The zero-curvature condition, therefore, gives
\eqn\zzz{-\phi_{uv} - {4\beta\over \alpha^2}\sum_{n = 1}^{\infty} n\gamma^n\sinh(2\beta n \phi) = 0.
}
We have, for $|\gamma| < 1$, the identity\foot{For finite $\phi$ it has a finite radius of convergence.}
\eqn\uuu{2\sum_{n = 1}^{\infty} n\gamma^n\sinh(n\phi) = {2\gamma(1 - \gamma^2)\sinh(\phi)\over \left[1 + \gamma^2 - 2\gamma \cosh(\phi)\right]^2} = {d\over d\phi}{\cal P}_\gamma(\phi).
}
This together with \zzz\ gives the equation of motion \aaax
\eqn\ttt{-\phi_{uv} -\alpha^{-2}\partial_\phi{\cal P}_\gamma(2\beta\phi) = 0.
}
This implies the monodromy $T(t, \lambda)$ exists and the theory has an infinite set of conserved charges (see the discussion in the previous section for details). For $\beta = 1/2$ and $\alpha = \sqrt{(1 - \gamma^2)/2}$ we get \aaaxx. We next compute the Poisson bracket for the Lax connection $x$ component $L$.

For the deformed sinh-Gordon model we have for the $x$ component of the connection
\eqn\ww{\eqalign{L & = L^a_\lambda\sigma_a = \beta \pi \sigma_0 +  {\beta\over \alpha}\sum_{n = 1}^{\infty}\left(\alpha_+^n(\phi)\sigma^+_n + \alpha^n_-(\phi)\sigma^-_n\right),
}
}
where $\sigma_a = \left\{\sigma_0, \sigma^{\pm}_n \right\}$ and
\eqn\xx{\pi = \partial_t \phi, \quad \alpha^n_{\pm}  = \gamma^{n/2}\left(\lambda e^{\pm n \beta\phi} + {1\over \lambda} e^{\mp n\beta\phi}\right).
}
The Poisson bracket is
\eqn\zz{\left\{L_1, L_2\right\} = \left\{L^a_\lambda(x), L^b_\mu(y)\right\}\left(\sigma_a\otimes \sigma_b\right).
}
We get, using the canonical equal-time relations,
\eqn\b{\eqalign{\left\{L_1, L_2\right\} & = \sum_{n = 1}^{\infty} \left[{\tilde\alpha}_+^n(\lambda) (\sigma^+_n\otimes \sigma_0) - {\tilde\alpha}_+^n(\mu)(\sigma_0\otimes \sigma^+_n) - {\tilde\alpha}_-^n(\lambda) (\sigma^-_n\otimes \sigma_0) + {\tilde\alpha}_-^n(\mu) (\sigma_0\otimes \sigma^-_n)\right]\delta(x-y),
}
}
where
\eqn\bbx{{\tilde\alpha}_{\pm}^n(\nu) =  {\beta^3 \over \alpha} n\gamma^{n/2}\left(\nu e^{\pm n \beta \phi(x)} - {1\over \nu}e^{\mp n \beta \phi(x)}\right).
}
Thus, it has the ultralocal or Sklyanin structure. This is sufficient for the charges to be in involution.

To identify the $r$-matrix, we look for a structure
\eqn\c{\left\{L_1, L_2\right\} = \left[r_{12}(\lambda, \mu), L_1(\lambda) + L_2(\mu)\right]\delta(x-y).
}
We write
\eqn\d{\eqalign{r_{12} & = r_{12}^{ab}\left(\sigma_a \otimes \sigma_b\right),\cr
& = r^{00}_{12}\left(\sigma_0\otimes \sigma_0\right) + r^{0+}_{12}\left(\sigma_0\otimes \sigma_+\right) + r^{+0}_{12}\left(\sigma_+\otimes \sigma_0\right) + r^{0-}_{12}\left(\sigma_0\otimes \sigma_-\right) + r^{-0}_{12}\left(\sigma_-\otimes \sigma_0\right)\cr
& + r^{++}_{12}\left(\sigma_+\otimes \sigma_+\right) + r^{--}_{12}\left(\sigma_-\otimes \sigma_-\right) + r^{-+}_{12}\left(\sigma_-\otimes \sigma_+\right) + r^{+-}_{12}\left(\sigma_+\otimes \sigma_-\right),\cr
& = r^{00}_{12}\left(\sigma_0\otimes \sigma_0\right) + \sum_{n = 1}^{\infty}r^{0n}_{12, 0+}\left(\sigma_0 \otimes \sigma^+_n\right) + \sum_{n = 1}^{\infty}r^{n0}_{12, +0}\left(\sigma^+_n \otimes \sigma_0\right) \cr
& + \sum_{n = 1}^{\infty}r^{0n}_{12, 0-}\left(\sigma_0 \otimes \sigma^-_n\right) + \sum_{n = 1}^{\infty}r^{n0}_{12, -0}\left(\sigma^-_n \otimes \sigma_0\right)\cr
& + \sum_{m, n = 1}^{\infty}r^{mn}_{12, ++}\left(\sigma^+_m \otimes \sigma^+_n\right) + \sum_{m, n = 1}^{\infty}r^{mn}_{12, --}\left(\sigma^-_m \otimes \sigma^-_n\right)\cr
& + \sum_{m, n = 1}^{\infty}r^{mn}_{12, +-}\left(\sigma^+_m \otimes \sigma^-_n\right) +\sum_{m, n = 1}^{\infty}r^{mn}_{12, -+}\left(\sigma^-_m \otimes \sigma^+_n\right).
}
}
Therefore,
\eqn\e{\eqalign{\left[r_{12}, L_1\right] & = r^{ab}_{12}L_\lambda^{c}\left[\sigma_a, \sigma_c\right]\otimes \sigma_b,\cr
\left[r_{12}, L_2\right] & = \sigma_a \otimes \left[\sigma_b, \sigma_c\right]r^{ab}_{12}L_\mu^{c}. 
}
}
We get
\eqn\g{\eqalign{r^{00}_{12} & = \beta^2{\lambda^2 +\mu^2\over \lambda^2 -\mu^2},\cr
r^{nk}_{12,+-} & = r^{kn}_{12,-+} = -\beta^2{2\lambda \mu\over \lambda^2-\mu^2}\delta_{nk}.
}
}
All the other components are zero. Thus, the $r$-matrix is constant and it is given by
\eqn\h{r_{12}(\lambda, \mu) = \beta^2\left\{{\lambda^2 +\mu^2\over \lambda^2 -\mu^2}(\sigma_0 \otimes \sigma_0) - {2\lambda \mu\over \lambda^2-\mu^2}\sum_{n = 1}^{\infty}(\sigma^-_n \otimes \sigma^+_n + \sigma^+_n \otimes \sigma^-_n)\right\} = -r_{21}(\mu, \lambda).
}
We note that for a given $n$ only $\sigma_0\otimes\sigma_0$, $\sigma^-_n\otimes\sigma^+_n$ and $\sigma^+_n\otimes\sigma^-_n$ appear. Moreover, the set $\sigma^-_n, \sigma_0$, $\sigma^+_n$ forms a Lie algebra. We now check the classical Yang-Baxter equation (CYBE) which is
\eqn\k{[r_{12}, r_{13}] + [r_{12}, r_{23}] + [r_{13}, r_{23}] = 0,
}
where
\eqn\l{
r_{12} = r^{ab}_{12}\left(\sigma_a\otimes \sigma_b \otimes I\right), \quad r_{13} = r^{ab}_{13}\left(\sigma_a\otimes I \otimes \sigma_b \right), \quad r_{23} = r^{ab}_{23}\left(I \otimes \sigma_a\otimes \sigma_b\right).
}
We note that for any given $n$
\eqn\n{ \eqalign{r^{00}_{12}r^{nn}_{13,+-} +  r^{nn}_{12,+-}r^{nn}_{23,+-} + r^{nn}_{13,+-}r^{00}_{23} & = 0,\cr
r^{nn}_{12,+-}r^{00}_{13} - r^{nn}_{12,+-}r^{00}_{23} - r^{nn}_{13,+-}r^{nn}_{23,-+} & = 0,\cr
 r^{nn}_{12,+-}r^{nn}_{13,-+} + r^{00}_{12}r^{nn}_{23,-+} - r^{00}_{13}r^{nn}_{23,-+} & = 0,\cr
}
}
and $r^{nn}_{ij, +-} = r^{nn}_{ij,-+}$. Thus, $[r_{12}, r_{13}] + [r_{12}, r_{23}] + [r_{13}, r_{23}] = 0$. Therefore, the Poisson bracket indeed satisfies the Jacobi identity. 

For the deformed sine-Gordon model \aaaz, the Lax connections are given by replacing $\phi$ with $i\phi$ in \yyy.

\newsec{Discussion} 

In this paper we considered the integrability of a deformed sine(sinh)-Gordon model. We showed that it is classically integrable. We also identified the underlying algebraic structure. The algebra \abb\ is a non-Lie alternative Akivis algebra. It is (loosely speaking) a direct sum of $sl(2, \IR)$ Lie algebras with a shared common Cartan. The Cartan acts on all the direct summands and thus the deformation (or extension) is neither a central extension nor an extension by derivation.\foot{One may refer to it as a hub-and-spoke deformation (or extension). It however gives, in general, an Akivis algebra which is not in general a Lie algebra.} It induces a grading by the adjoint action. See Fig. 1 for a pictorial representation of the algebra structure.

To the best of my knowledge the algebra \abb\ is new and it has not been reported elsewhere in the literature. It supports integrable field theories although it is a non-Lie and non-associative algebra.

We also note that \zzz\ with finite terms in the sum,  
\eqn\zzzs{-\phi_{uv} - {4\beta\over \alpha^2}\sum_{n = 1}^{N} n\gamma^n\sinh(2\beta n \phi) = 0,
}
corresponds to an integrable model. Therefore, we actually have a family of infinite classically integrable models. The underlying algebra is a ${\cal Z}_{2N + 1}$-graded non-Lie alternative Akivis algebra. It satisfies the identities \newaeeb\ and \newaee. In Fig. 1, for example, $N = 4$. For $N = \infty$, it reduces to \aaax\ which we write again here for later convenience, 
\eqn\zzzx{-\phi_{uv} - {4\beta\over \alpha^2}{\gamma(1 - \gamma^2)\sinh(2\beta\phi)\over(1 + \gamma^2 - 2\gamma\cosh(2\beta\phi))^2} = 0.
}
In the double scaling limit \AsratB, in which we take $\gamma \to 0$ and $\phi \to \infty$ but we keep the combination $\gamma e^{2\beta\phi}$ finite and fixed, we find, for finite $N$, the integrable model
\eqn\zzzsccc{-\phi_{uv} - {2\beta\over \alpha^2}\sum_{n = 1}^{N} n\gamma^n e^{2\beta n \phi} = 0.
}
For $N = \infty$, this simplifies to
\eqn\zzzxxcc{-\phi_{uv} - {2\beta\over \alpha^2}{\gamma e^{2\beta\phi}\over(1 - \gamma e^{2\beta\phi})^2} = 0.
}
It describes a deformed Liouville theory.

The analytic continuation $\phi \to i\phi$ in \zzzs\ gives
\eqn\zzzsr{-\phi_{uv} - {4\beta\over \alpha^2}\sum_{n = 1}^{N} n\gamma^n\sin(2\beta n \phi) = 0.
}
This is also classically integrable. For $N = \infty$, it simplifies to \aaaz,
\eqn\zzzxx{-\phi_{uv} - {4\beta\over \alpha^2}{\gamma(1 - \gamma^2)\sin(2\beta\phi)\over(1 + \gamma^2 - 2\gamma\cos(2\beta\phi))^2} = 0.
} 
Note here $\gamma$ can be positive, therefore, we take $|\gamma| < 1$.

We also note that \zzzs\ can be put into the form
\eqn\zzzss{\eqalign{-\phi_{uv} & = {2\beta\over \alpha^2}\sum_{n = -N}^{N} n\gamma^{|n|} e^{2\beta n\phi},\cr
& = -{1\over \varepsilon}\sum_{n = -N}^{N} c_n^0q_n e^{-q_n\phi} , \quad q_n = -2\beta n, \quad \varepsilon = \alpha^2, \quad c^0_n = \gamma^{|n|}.
}
}
For $0 < \gamma < 1$, this equation is also known as the Poisson-Boltzmann equation. It describes, in thermal equilibrium, how mobile ions or solutes redistribute around charged surfaces in a medium or a solvent. For positive $n$, $q_n$ is the charge for an anion, and for negative $n$, $q_n$ is the charge for a cation. $c^0_n$ is the mean or bulk concentration of ion $n$. $\varepsilon$ is the permittivity of the medium or solvent, and it is always greater than one. $\phi$ is the electric potential. Note that 
\eqn\zzzr{\sum_{n = -N}^{N} c_n^0q_n = 0.
} 
This is the bulk electroneutrality condition. This ensures that the system is stable.

For finite $N$ the problem is well defined. However, for $N = \infty$, as we can see from \zzzx, $\phi$ must satisfy 
\eqn\zzzt{|\phi| < \phi_0 = -{1\over 2\beta}\ln(\gamma),
}
$\phi_0$ is the maximum value of the potential. Understanding what happens when $\phi \geq \phi_0$ could give a physical insight on (the fate of) the curvature singularity in the string background ${\cal A}_3$ in the case $\gamma$ is positive. And in general on spacetime singularities.

To the best of my knowledge the integrable models \zzzs\ and \zzzsr\ are new and they have not been reported elsewhere in the literature before. The models are interesting and special because they are based on non-Lie algebras. Moreover, the algebras are non-associative. We expect, more generally, that the family of integrable models with non-Lie underlying algebras to be broad. 

In this paper we did not compute explicitly the conserved quantities. We also did not consider the models quantum integrability. We hope to address these and related questions in a separate paper in the future. We also hope to extend the hub-and-spoke construction to other Lie algebras.

\bigskip\bigskip

\noindent{\bf Acknowledgements:} I thank ChatGPT and Gemini for useful literature search and information.

\appendix{A}{The flatness condition}
We show that the auxiliary linear problem \oozxzoo\ is equivalent to the zero-curvature condition \cc. We have
\eqn\compzz{\eqalign{ (\partial_x\partial_t)\psi &  = \partial_x(\partial_t\psi),\cr
& = \partial_x(M\psi),\cr
& = (\partial_x M) \psi + M (\partial_x\psi),\cr
& = \partial_x M \psi + M(L\psi).
}
}
Similarly, we obtain
\eqn\compzzz{(\partial_x\partial_t)\psi  = \partial_t(\partial_x\psi) = \partial_t L \psi + L(M\psi).
}
Together, \compzz\ and \compzzz, gives
\eqn\compzzzz{\partial_t L \psi - \partial_x M \psi + L(M\psi) - M(L\psi) = 0.
}
This gives the flatness or zero-curvature condition \cc. Note the multiplication is in the universal enveloping algebra.
 
\appendix{B}{The Monodromy matrix}
The monodromy matrix (or more generally the state-transition matrix) is defined as
\eqn\zzccx{
T(x, y, \lambda) = {\cal P} e^{\int_{y}^{x} dx'L(x', \lambda)},
}
where ${\cal P}$ denotes a path ordering (and a particular nesting prescription of parentheses), and $L$ is the spatial component of the Lax connection. $T$ satisfies the relations
\eqn\wzzaax{\eqalign{
T(x, x, \lambda) & = 1 := e,\cr
T(y, x, \lambda) & = T^{-1}(x, y, \lambda),\cr
T(x, y, \lambda) & = T(x, z, \lambda)T(z, y, \lambda),\cr
\partial_x T(x, y, \lambda) & = L(x, \lambda)T(x, y, \lambda),\quad ({\rm in \ matrix \ algebra})\cr
& = \rho_T L(x, \lambda), \quad ({\rm in\ general})\cr
\partial_y T(x, y, \lambda) & = - T(x, y, \lambda) L(y, \lambda),\quad ({\rm in \ matrix \ algebra})\cr
& = -\lambda_T L(y, \lambda), \quad ({\rm in\ general})
}
}
where $e$ denotes the identity and $x > z > y$. Using the zero-curvature condition \compzzzz\ (or \cc), we have
\eqn\zzaax{\eqalign{
\partial_t T(x, y, \lambda) & = {\cal P} \int_y^x dz  e^{\int_z^x dx' L(x', \lambda)} \partial_t L(z, \lambda) e^{\int_y^z dx' L(x', \lambda)},\cr
& = \int_y^x dz T(x, z, \lambda)\partial_t L(z, \lambda) T(z, y, \lambda),\cr
& = \int_y^x dz T(x, z, \lambda)\left(\partial_z M(z, \lambda) - [L(z, \lambda), M(z, \lambda)]\right)T(z, y, \lambda),\cr
& = \int_y^x dz \partial_z \left( T(x, z, \lambda)M(z, \lambda) T(z, y, \lambda)\right),\cr
\partial_t T(x, y, \lambda) & = M(x, \lambda)T(x, y, \lambda) - T(x, y, \lambda)M(y, \lambda),\quad ({\rm in \ matrix \ algebra})\cr
& = \rho_T M(x, \lambda) - \lambda_T M(y, \lambda). \quad ({\rm in\ general})
}
}
The relations \wzzaax\ and \zzaax\ can be alternatively taken as the definition of the monodromy matrix (or more generally the state-transition matrix). Under periodic boundary conditions we have $M(x=0, \lambda) = M(x=2\pi, \lambda) = M(\lambda)$. This gives in general \ttm,
\eqn\appzz{{\partial T(\lambda)\over \partial t} = \rho_T M(\lambda) - \lambda_T M(\lambda),
}
where $T(\lambda) = T(2\pi, 0, \lambda)$. When the spatial coordinate is on a line, it is the infinite volume limit of the state-transition matrix $T(x, y, \lambda)$, \ie, $T(\infty, -\infty, \lambda) := T(\lambda)$, with proper regularization, that gives the infinite conserved charges.

The Poisson bracket of $T_1 = T \otimes I$ and $T_2 = I\otimes T$ is given by
\eqn\zzbbx{\eqalign{
\{T_1(x, y, \lambda), T_2(x, y, \mu)\} & = \int_y^x \int_y^x  dz dw T_1(x, z, \lambda)T_2(x, w, \mu) \{L_1(z, \lambda), L_2(w, \mu)\} T_1(z, y, \lambda) T_2(w, y, \mu), \cr
& = \int_y^x  dz T_1(x, z, \lambda)T_2(x, z, \mu) [r_{12}, L_1(z, \lambda) + L_2(z, \mu)] T_1(z, y, \lambda) T_2(z, y, \mu),\cr
& =  \int_y^x  dz \partial_z\left(T_1(x, z, \lambda)T_2(x, z, \mu) r_{12}T_1(z, y, \lambda) T_2(z, y, \mu)\right),\cr
\{T_1(x, y, \lambda), T_2(x, y, \mu)\} & = r_{12}T_1(x, y, \lambda)T_2(x, y, \mu) - T_1(x, y, \lambda)T_2(x, y, \mu)r_{12}.\quad ({\rm in \ matrix \ algebra})
}
}
We have used the Sklyanin bracket, see \eg\ \llzxzll, and thus the $r$-matrix is constant. In terms of the differentials of the right and left translations \ttm\ we have in general (for periodic boundary conditions)
\eqn\zzddx{
\{T_1(\lambda), T_2(\mu)\} = \rho_{T_1}\rho_{T_2}r_{12}(\lambda, \mu) - \lambda_{T_1}\lambda_{T_2}r_{12}(\lambda, \mu), \quad ({\rm in \  general})
}
where $T_n(\nu) = T_n(2\pi, 0, \nu)$, and $\rho_{T_1} = \rho_T\otimes I, \rho_{T_2} = I \otimes \rho_T, \lambda_{T_1} = \lambda_T\otimes I, \lambda_{T_2} = I \otimes \lambda_T$.

\listrefs

\bye